# Deposition of MAX phase containing thin films from a (Ti,Zr)$_2$AlC compound target

Clio Azina[a,*,1], Bensu Tunca[b,c] Andrejs Petruhins[a], Binbin Xin[a], Melike Yildizhan[a], Per O.Å. Persson[a], Jozef Vleugels[b], Konstantina Lambrinou[c,d], Johanna Rosén[a], Per Eklund[a]

[a] Thin Film Physics Division, Department of Physics (IFM), Linköping University, SE-581 83, Linköping, Sweden

[b] KU Leuven, Department of Materials Engineering, Kasteelpark Arenberg 44, B-3001 Leuven, Belgium

[c] SCK CEN, Boeretang 200, B-2400 Mol, Belgium

[d] School of Computing and Engineering, University of Huddersfield, Queensgate, Huddersfield HD1 3DH, UK

* Corresponding author. E-mail address: clio.azina@liu.se (C. Azina)

---

[1] Current affiliation: Materials Chemistry, RWTH Aachen University, Kopernikusstr. 10, D-52074, Aachen, Germany; azina@mch.rwth-aachen.de

**Abstract**

This work reports on sputter depositions carried out from a compound $(Ti,Zr)_2AlC$ target, whereupon Al-containing $(Ti,Zr)C$ thin films (30-40 nm in thickness) were deposited on MgO(111) and $Al_2O_3$(0001) substrates at temperatures ranging between 500 and 900 °C. The presence of Al within the carbide structure was evidenced by lattice parameter variations. Furthermore, chemical analyses showed that the Al distribution throughout the film thickness was fairly homogeneous. Thicker films (80-90 nm) deposited from the same compound target consisted of the pseudo-binary $(Ti,Zr)C$ and intermetallic compounds in the Ti-Zr-Al system up to 800 °C, as well as solid solution MAX phases with different Ti:Zr ratios at 900 °C. X-ray diffraction and transmission electron microscopy showed that both $(Ti,Zr)_2AlC$ and $(Ti,Zr)_3AlC_2$ solid solution MAX phases were formed. Moreover, this work discusses the growth mechanism of the thicker films, which started with the formation of the mixed $(Ti,Zr)C$ carbide, followed by the nucleation and growth of aluminides, eventually leading to the formation of the MAX phases, which was the primary objective of the sputter depositions.

# 1. Introduction

Binary carbides based on early transition metals are an important class of materials in applications involving cutting and drilling tools because of their high hardness [1–4]. Beyond the traditional binary carbides, more phases exist, including the inherently laminated ternary carbides and nitrides known as MAX phases. MAX phases are described by the general formula $M_{n+1}AX_n$, where M is an early transition metal, A is an A-group element, X is carbon and/or nitrogen and n = 1, 2 or 3. The MAX phases are gaining a lot of attention because of their exceptional ceramic/metal hybrid properties and the broad spectrum of potential applications [5–8]. Due to their structural complexity, a large Gibbs free energy barrier is often associated with the synthesis of MAX phase compounds; therefore, complex pseudo-binary phases may precede the nucleation of MAX phase compounds, making their formation an energetically more favourable process [9–11]. The formation of a MAX phase compound can thus be affected by the presence of pseudo-binary phases that enable the growth of the particular MAX phase by diffusion-controlled processes [9,11–14].

While some MAX phases (e.g., $Cr_2AlC$ [15,16]) have been directly deposited without the need of a buffer layer at lower temperatures (~500 °C), MAX phase compounds in the Ti-Al-C system are usually deposited at higher temperatures and often require either a $TiC_x$ buffer layer and/or an annealing step [17–20]. For instance, Wilhelmsson et al. discussed the deposition of ternary films in the Ti-Al-C system, using a $TiC_x$ seed layer [20]. These authors showed that even though MAX phase compounds in that system are obtained at high temperatures (> 800 °C), Al starts dissolving in the $TiC_x$ seed layer at temperatures as low as 300 °C. Upon the physical vapour deposition (PVD) of thin films, non-equilibrium phases can readily occur. For example, the concentration of dissolved Al in Ti-Al-C thin films can be much higher than the equilibrium concentration [20]. Furthermore, Frodelius et al. reported the deposition of a (Ti,Al)C carbide solid solution, when attempting to deposit the $Ti_2AlC$ MAX phase from the respective target [19].

It has also been shown that it is possible to tune the properties of the MAX phases by tailoring their chemistry. In fact, MAX phase solid solutions have been the subject of numerous studies, irrespective of whether the solid solution was formed on the M, A or X site [21–29]. Depending on the considered alloying element and the exact solid solution stoichiometry one may achieve specific functionalities and tailor the material properties. For example, Cabioc'h et al. reported on the tailoring of the thermal expansion coefficient of $Cr_2(Al_x,Ge_{1-x})C$ MAX phases, showing a clear evolution with respect to the A-metal ratio [29]. Another example worth mentioning is that of magnetic MAX phases, where magnetism is caused by the introduction of Mn [26,30]. Sputter deposition from elemental targets is typically preferred for the investigation of growth processes, due to the more precise control of the thin film composition. However, industrial material development requires simpler and faster coating processes, which includes the use of compound targets. While the use of a single target simplifies the deposition process, the obtained film composition is very often different than that of the target, as previously discussed for $Ti_3SiC_2$ [31,32], $Ti_2AlC$ [19] and $Cr_2AlC$ [33].

The Ti-Zr-Al-C system was selected in this work, as MAX phase compounds in this system are promising candidate coating materials for accident-tolerant fuel (ATF) cladding materials intended for Gen-II/III light water reactors (LWRs), particularly due to the low neutron cross-section of zirconium (Zr). Most prior studies on Zr-based MAX phases reported the presence of secondary phases (i.e., binary carbides and intermetallics), particularly in bulk ceramics [34,35]. In the case of thin films, the MAX phases have not been previously obtained, however, the formation of laminated carbides, such as $Zr_2Al_3C_4$ and $Zr_2Al_4C_5$, has been reported [36,37]. This study investigates the binary carbides and ternary carbides (MAX phases) formed when growing films from a MAX phase target by means of sputter deposition at various substrate temperatures. The initial growth of (Ti,Zr)C pseudo-binary carbides was observed after short deposition times of 30 min. Longer deposition times at 900 °C resulted in thin films containing more than one phase, including the targeted $(Ti,Zr)_2AlC$ MAX phase solid solution.

## 2. Materials and Methods

ZrH$_2$ (<6 μm, >99% purity, Chemetall, Germany), TiH$_2$ (<8 μm; >99% purity; Chemetall, Germany), Al (<5 μm, >99% purity, AEE, USA), and C (<5 μm, >99% purity, Asbury Graphite Mills, USA) were used as starting powders for the synthesis of the MAX phase target. For this purpose, TiH$_2$:Zr:Al:C powders were mixed in isopropanol for 24 h in argon (Ar) in an atomic ratio of 1.5:1.5:1.2:1.9, which is the starting composition of a C-substoichiometric (Ti,Zr)$_3$AlC$_2$ MAX phase (312). Due to parasitic reactions that result mainly in carbide formation, a 312 starting composition produces the (Ti,Zr)$_2$AlC MAX phase (211) and pseudo-binary carbides [34]. The powders were reactively hot pressed (W100/150-2200-50 LAX, FCT Systeme, Frankenblick, Germany) in vacuum (~10 Pa), in a graphite die setup (∅ 56 mm), at 1450°C for 1 h and under 30 MPa pressure. A 0.5 mm-thick material layer that included the outer carbides was removed from the top and bottom surfaces of the sintered disc by grinding, and the disc was reduced to a 2-inch diameter target, using electrical discharge machining (EDM). This target will herein be referred to as (Ti,Zr)$_2$AlC, since its main component was the 211 MAX phase compound, even though the composition of the starting powder mixture corresponded to the 312 MAX phase compound.

Thin films were deposited by magnetron sputtering using the ∅ 50 mm (2-inch) (Ti,Zr)$_2$AlC compound target. The base pressure in the deposition chamber was kept below 2.6×10$^{-5}$ Pa, while the working pressure in the presence of Ar was 0.6 Pa. The target was placed on-axis, at a distance of 180 mm below the substrate holder. The target was operated in the constant power mode at 50 W, and the substrates were rotated at 10 rpm. Thin film depositions were performed at different temperatures in the 500-900 °C range, at 100 °C increments. All depositions lasted for 30 min on MgO(111) and Al$_2$O$_3$(0001) substrates that were cleaned ultrasonically prior to deposition in acetone and isopropanol, for 5 min in each medium. These depositions resulted in thin films with 30-40 nm in thickness. Depositions of 60 and 90 min were carried out only on Al$_2$O$_3$(0001), resulting in films of 80-90 nm and ~120 nm in thickness, respectively.

The structural properties of the target and the deposited films were investigated by means of X-ray diffraction (XRD), using a standard *θ-2θ* geometry in a Panalytical X'pert MRD with Cu $K_α$ radiation. Film thicknesses were obtained using X-ray reflectivity (XRR) in a Philips X'Pert MRD system and were confirmed by scanning electron microscopy (SEM; Zeiss Leo 1550 Gemini).

Time-of-flight-energy elastic recoil detection analysis (ToF-E ERDA) was performed at the Tandem Accelerator Laboratory of Uppsala University. This technique used 36 MeV $^{127}I^{8+}$ ions as incident beam impinging onto the samples at an incoming angle of 67.5 ° with respect to the surface normal, and the induced recoiling particles were detected by two carbon foil time detectors and a silicon p-i-n diode energy detector for energy discrimination at an angle of 45° with respect to the incoming ion beam. The measured ERDA spectra were converted to depth profiles using the Potku code [38].

Scanning transmission electron microscopy (STEM) analyses were performed using a JEOL ARM200F Cs-corrected scanning transmission electron microscope (S/TEM), operated at 200 kV. Low magnification high-angle annular dark-field (HAADF) STEM images, as well as high resolution (HRTEM) images were acquired. Energy dispersive X-ray spectroscopy (EDS) was conducted in the STEM mode. Selected area electron diffraction (SAED) patterns of the film and the interface were acquired and fast Fourier transforms (FFT) were performed on the HRTEM micrographs. Thin foils were lifted out from the areas of interest, using a Carl Zeiss Cross-Beam 1540 EsB focused ion beam (FIB).

3. **Results and Discussion**

The $(Ti,Zr)_2AlC$ target was analysed prior to thin film deposition, in order to verify its phase assembly and compositional homogeneity. Fig. 1a shows a schematic view of the analysed piece of a dimensionally smaller, but otherwise equivalent, MAX phase target. Fig. 1b to 1d shows results of the SEM/EDS analysis on different target regions, which confirm the presence of different phases. Along the target radius and close to the target outer surface (zone C), two types of 211 MAX phases formed, one with a 50:50 Ti:Zr ratio (green squares) and one with a 70:30 Ti:Zr ratio (red circles). Towards the centre of the target (zones A and B), more of the 211 phase with a 70:30 Ti:Zr ratio was found. $Al_2Zr$ and (Ti,Zr)C have also been detected across the target thickness. The mixed (pseudo-binary) carbides appeared to be primarily Zr-rich, although some Ti-rich particles have also been observed, mainly due to the expected carbide decomposition under the selected synthesis conditions [28]. Fig. 1e shows that the elemental distribution did not vary significantly across the target diameter; therefore, the target chemical composition was considered to be homogeneous, unlike the phase distribution that varied across the target. The XRD pattern of the target before sputtering shows peaks of the MAX phase as well as peaks of the pseudo-binary aluminides and carbides (Fig. 1f). While the aluminides and carbides were Zr-rich, their peaks were slightly shifted due to the presence of dissolved Ti in the crystal lattice.

The target was used to deposit thin films of variable thickness on MgO and $Al_2O_3$ substrates. Fig. 2 shows the XRD $\theta$-$2\theta$ patterns from the films deposited for 30 min on both $Al_2O_3$ and MgO substrates, at temperatures ranging between 500 and 900 °C. X-ray reflectivity was used to determine the thickness of the produced thin films, which varied in the 30-40 nm range for 30 min-long depositions. The pseudo-binary carbide $(Ti_x,Zr_{1-x})C$ was identified by the XRD peaks at ~34.3° and ~73° $2\theta$. The intensity of the pseudo-binary carbide XRD peaks increased with increasing deposition temperature, which may be associated with the presence of larger grains. The comparatively lower nucleation temperature of (Ti,Zr)C on MgO may be the result of the milder interaction between carbide and substrate. In the case of the $Al_2O_3$ substrate, diffusion processes that occurred at the coating/substrate interface affected the nucleation of the carbide, requiring higher temperatures to stabilize this carbide phase.

The compositions and lattice parameters of the carbide thin films deposited on MgO(111) substrates are presented in Table 1. The lattice parameters of the carbides did not follow a particular trend as function of the substrate temperature. Therefore, one can assume that these differences are related to the Al content in the carbide, which affects the lattice parameter.

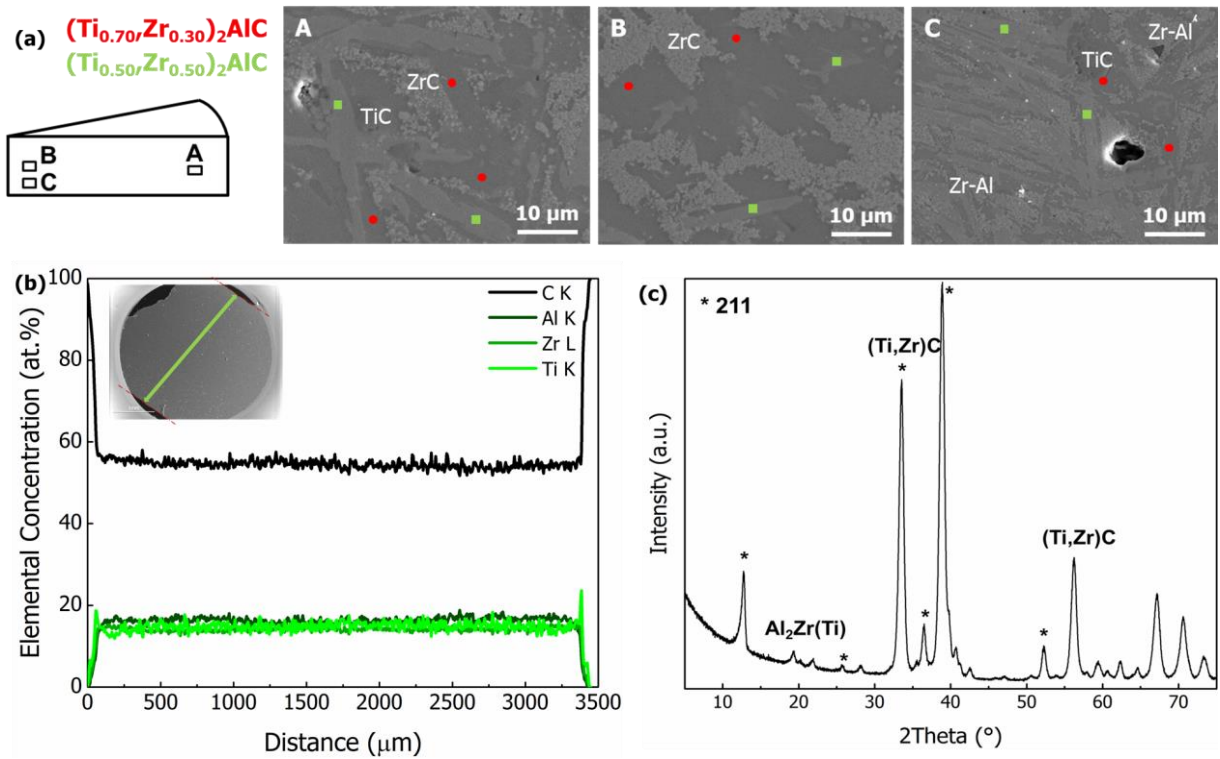

**Fig. 1.** (a) Schematic representation of the analysed cross-section through an equivalent MAX phase target with smaller diameter, and corresponding SEM micrographs and phases detected in the target (A-C). (b) EDS line scans across the target. (c) XRD pattern of the target before sputter deposition.

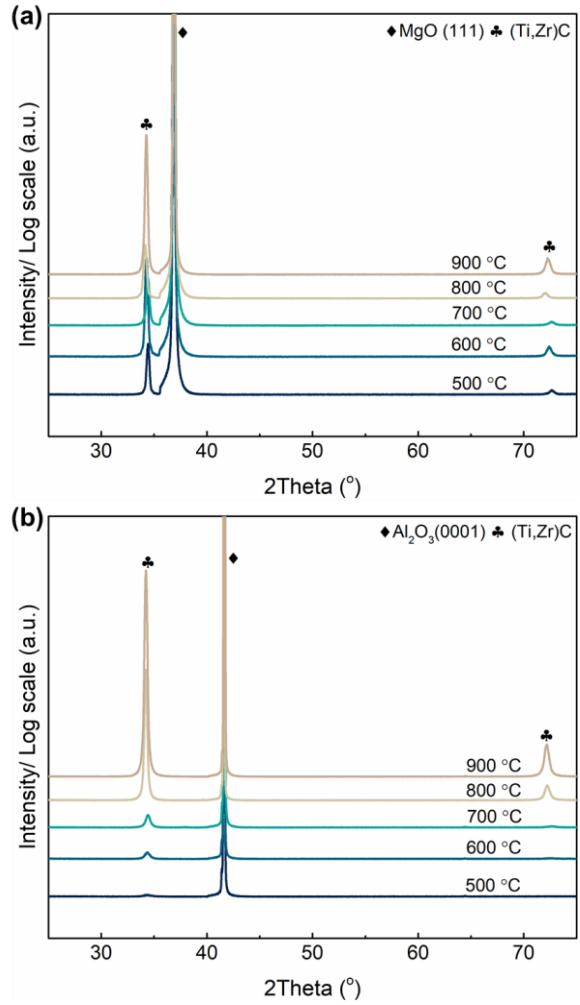

**Fig. 2.** XRD patterns of thin films deposited in 30 min from the (Ti,Zr)$_2$AlC compound target on (a) MgO(111) and (b) Al$_2$O$_3$(0001) substrates.

**Table 1.** Lattice parameters and composition of thin films deposited in 30 min at different temperatures on MgO (111) substrates.

| Temperature (°C) | Lattice parameter $a$ (Å) | Ti (at.%) | Zr (at.%) | Al (at.%) | C (at.%) |
|---|---|---|---|---|---|
| 500 | 4.50470 | 26 ± 1 | 25 ± 1 | 17 ± 1 | 32 ± 1 |
| 600 | 4.51788 | 26 ± 1 | 23 ± 1 | 14 ± 1 | 37 ± 1 |
| 700 | 4.50723 | 27 ± 1 | 25 ± 1 | 14 ± 1 | 34 ± 1 |
| 800 | 4.53135 | 29 ± 1 | 29 ± 1 | 9 ± 1 | 33 ± 1 |
| 900 | 4.52366 | 28 ± 1 | 26 ± 1 | 12 ± 1 | 34 ± 1 |

The determined by ERDA compositions of thin films deposited at different temperatures on MgO substrates are given in **Table 1**. MgO substrates were used to be able to extract the Al contribution. Fig. 3 shows two representative compositional depth profiles obtained from the films deposited at 800 and 900 °C in 30 min. The different peak shapes relate to differences in thin film thickness and density: the film deposited at 900 °C was thicker, but less dense than the one deposited at 800 °C. The relative concentrations were acquired from the total thin film thickness. The Ti:Zr ratios were very close to 1:1 and corresponded to the respective elemental ratio in the target. The C composition appeared unaffected by the deposition temperature and was slightly higher than that of a 2:1:1 MAX phase composition, as expected from the C content of the pristine target that corresponded to the 312 MAX phase. Moreover, the thin film contained 9-17 at.% Al, depending on the deposition temperature. It is also worthwhile noticing that the Al distribution was fairly homogeneous over the entire thin film thickness.

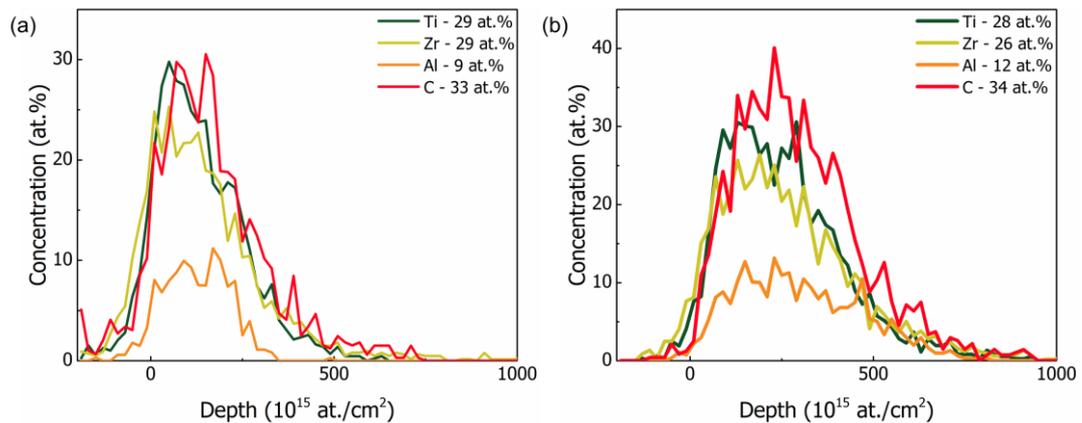

**Fig. 3.** Compositional depth profiles of thin films deposited on MgO (111) substrates at (a) 800 and (b) 900 °C.

In order to obtain thicker films for further analysis, 60 min-long depositions were carried out, which resulted in 80-90 nm-thick films. XRD patterns collected from these films are shown in Fig. 4. One may notice that the $(Ti_x,Zr_{1-x})C$ pseudo-binary carbide is present and appears to be the only constituent phase of the films deposited at 500 and 600 °C. At 700 and 800 °C, peaks appeared at $2\theta$ ~38° and ~45° that corresponded to an $Al_3(Zr,Ti)$ intermetallic phase. At 900 °C, the $Al_2Zr$ intermetallic was indexed with (002) and (006) reflections at ~20° and ~64.5°, respectively. The Al$_2$Zr peaks were slightly shifted towards higher 2$\theta$ angles, indicating Zr-substitution by Ti, i.e., Al$_2$(Zr,Ti) formation. A 211 MAX phase was also formed at 900 °C, as indicated by the peaks at 2$\theta$~12.8° (002), ~25.8° (004) and ~39.2° (006). However, these peaks fit quite poorly with the peak positions reported by Tunca et al. for a Ti:Zr ratio of 1:1 [28]. The mismatch is mostly due to internal stresses, as will be shown later. Furthermore, a clear peak shift was observed for the (Ti,Zr)C pseudo-binary carbides towards lower angles, indicating that this carbide was Zr-rich at higher temperatures.

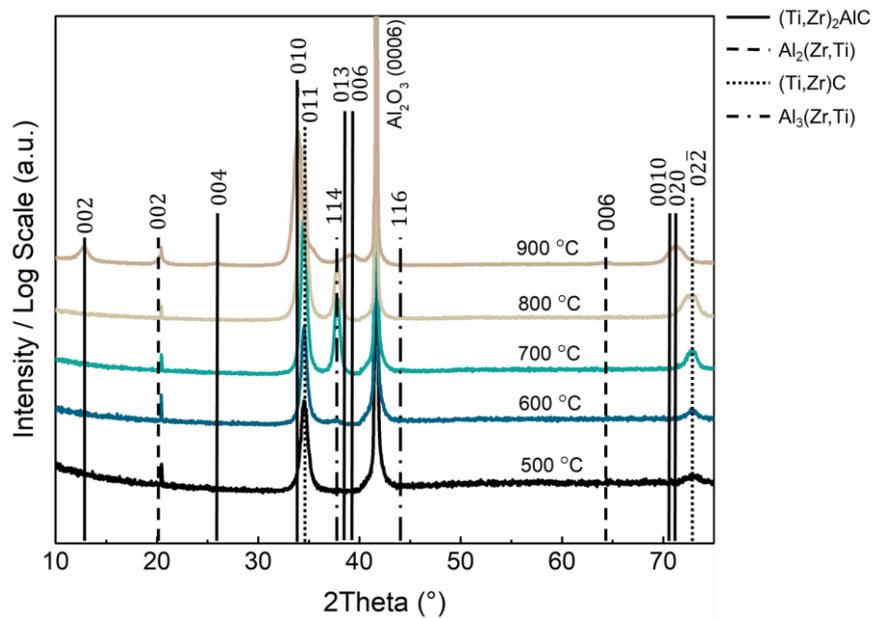

**Fig. 4.** XRD patterns of thin films deposited in 60 min from the (Ti,Zr)$_2$AlC compound target on Al$_2$O$_3$(0001) substrates.

The formation of a Ti-rich MAX phase is herein addressed by analysing the sequence of MAX phase formation. As intermetallic compounds are often involved in the formation of various MAX phases, the following sequence of phase formation may be proposed: first, a (Ti,Zr)C pseudo-binary carbide formed, followed by the formation of (Ti,Zr)Al intermetallics (IMCs) that, at higher temperatures, reacted with the existing carbide to form the MAX phase. Furthermore, the reaction between intermetallic and carbide favoured the depletion of the latter in Ti rather than in Zr, as ZrC is a more stable compound, thus promoting the formation of a Ti-rich MAX phase compound. The synthesis of this solid solution MAX phase was experimentally achieved, however, a sufficiently thick carbide interlayer was required, as deduced from the comparison of the 30 min-long and 60 min-long depositions. Magnuson et al. have reported that TiC formed as result of the excess carbon in a $Ti_3SiC_2$ target [39]. Moreover, a substoichiometric carbide interlayer has been found to undergo a solid state reaction with the $Al_2O_3$ substrate at elevated temperatures to form the MAX phase at the interface. [40,41] Furthermore, one can assume that to form a phase pure MAX phase film, the substrate temperature should above 900 °C. Such high temperatures are, however, not recommended when considering these coatings for ATF cladding materials, as the intended zircaloy substrates should not be exposed to temperatures higher than 600 °C.

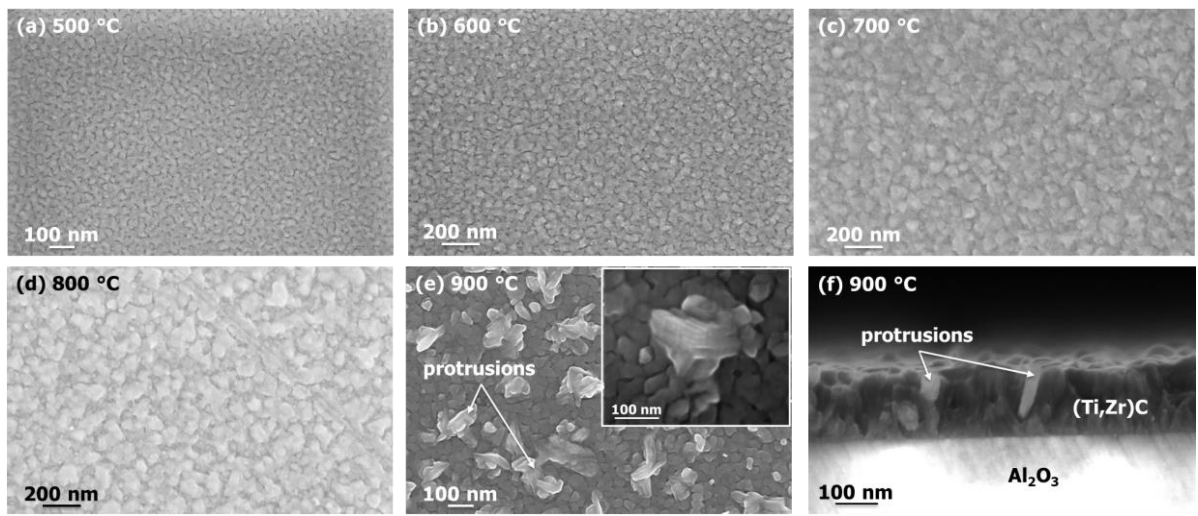

**Fig. 5.** SEM micrographs of the surface of the 80 nm-thick films deposited in 60 min on $Al_2O_3$ (0001) substrates at (a) 500, (b) 600, (c) 700, (d) 800, and (e) 900 °C. (f) SEM image of a cross-section of a thin film deposited on $Al_2O_3$ (0001) substrate at 900 °C in 90 min.

Representative SEM micrographs of the produced thin films in Fig. 5 show the increase in grain size when the deposition temperature increased from 500 to 800 °C. At 900 °C, the presence of at least two phases and two different grain morphologies may be observed. The brighter grain agglomerates that protrude from the thin film surface, thereby increasing the thin film roughness, are stackings of ultra-thin (<30 nm) laths (Fig. 5e, magnified inset). The darker grains correspond to a typical carbide microstructure composed by randomly oriented, equiaxed grains. TEM revealed that the protruding grain stackings correspond to the MAX phase, while the rest of the film is composed of (Ti,Zr)C carbide grains. Fig. 5f shows the cross-section of a ~160 nm-thick film, the microstructure of which is inhomogeneous across the film thickness. Closer to the substrate, small grains are observed. Towards the film surface, the film displays columnar growth.

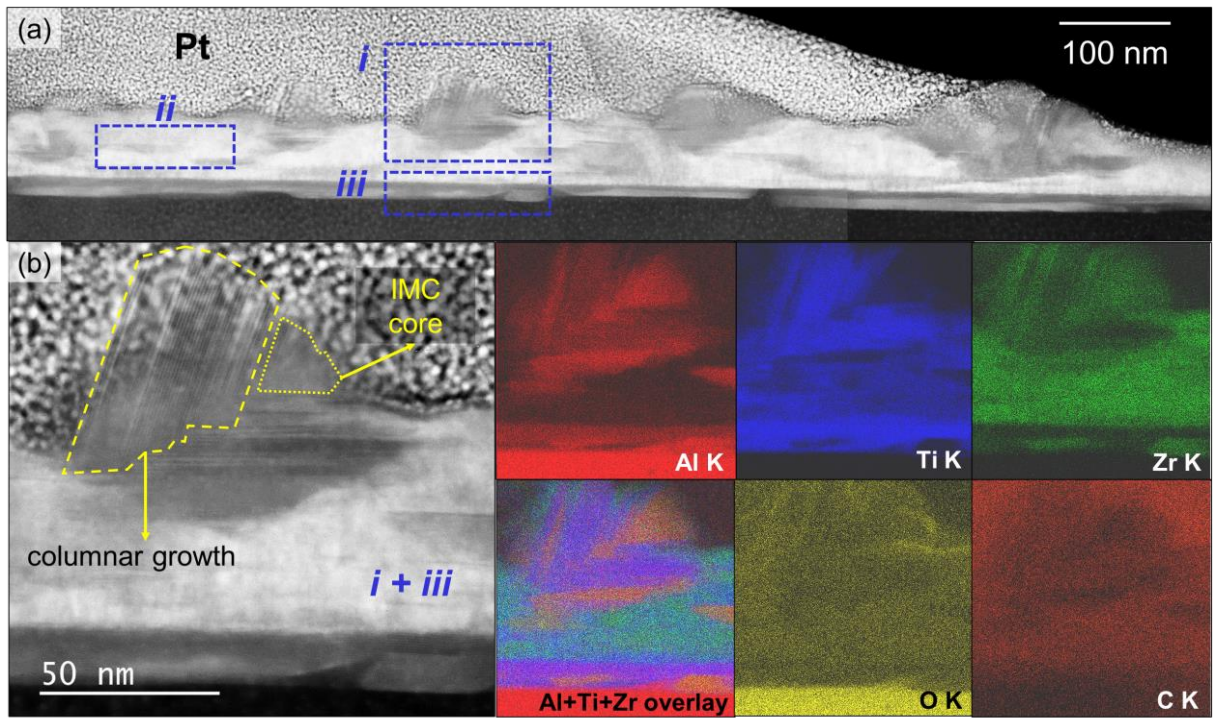

**Fig. 6**. (a) Overview of the thin film deposited at 900 °C in 90 min on an $Al_2O_3$ substrate: columnar grain growth (area *(i)*), continuous film bulk (area *(ii)*), and film/substrate interface (area *(iii)*). (b) Columnar grain growth and IMC core observed in areas *(i)* and *(iii)*; Al, Ti, Zr, O, C EDS elemental maps of this region in STEM mode and Al+Ti+Zr map overlay.

STEM analysis of a FIB foil extracted from the thin film deposited at 900 °C in 90 min on an $Al_2O_3$ substrate is shown in Fig. 6. A low-magnification HAADF STEM image of the thin film is shown in Fig. 6a. At first glance, the thin film is characterised by three distinct microstructural features: (a) the darker, columnar features (area *(i)*) corresponding to the grain protrusions shown in Fig. 5e, (b) the brighter, continuous film (area *(ii)*), and (c) the film/substrate interface (area *(iii)*). Detailed compositional analysis of these features revealed that the continuous film in the STEM images contained more Zr, while neither the continuous film nor the film/substrate interface were compositionally homogeneous. The grain protrusions in Fig. 5e were composed of a faceted core (dotted area, Fig. 6b) surrounded by a columnar-like structure (dashed area, Fig. 6b). Both microstructural features are also visible in the STEM image and elemental maps of Fig. 6b. The cores were Al-rich, whereas the columnar ones were transition metal-rich. In terms of oxygen content, there is slightly more O at the thin film surface and at the interface with the $Al_2O_3$ substrate; however, there is no specific trend in O level variation as function of coating thickness. The overall O content of the thin film was somewhere between 9 and 13 at.%.

In order to understand the crystal structure of the different microstructural features, high magnification STEM micrographs and FFT patterns of both the thin film and the film/substrate interface were acquired. Fig. 7 shows a high magnification STEM micrograph from area *(i)* in Fig. 6a, corresponding to the protrusions observed in Fig. 5e. As shown in Fig. 7a, the columnar structures correspond to mixed 211 and 312 (Fig. 7b) MAX phase atomic stackings. They seem to nucleate on the Al-rich core, which was indexed as $Al_3(Zr,Ti)_2$ (Fig. 7c). It appears that the intermetallic provides Al to the (Ti,Zr)C carbide, leading to a 211 Al-rich MAX phase parallel to the (111) planes of the intermetallic core. Further away from the IMC core, the 211 MAX phase becomes 312, where contains less Al. A sequence of 211 to 312 and inversely can be observed in Fig. 7a. Such features are found both within the film thickness and closer to its surface. However, the lack of 312 phase contributions in the XRD patterns in Fig. 4 indicates that the 312 phase is only present in small amounts.

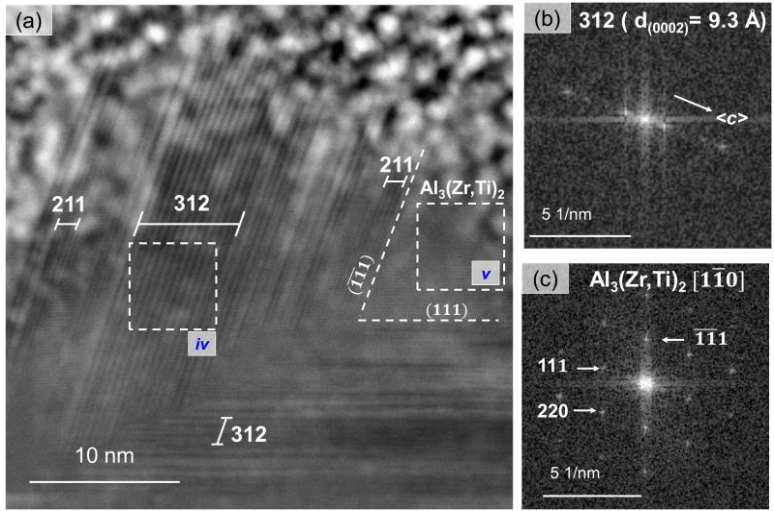

**Fig. 7**. (a) High magnification STEM micrograph of a protrusion showing the columnar MAX phase and IMC features. (b) FFT pattern of area *(iv)* indicative of the 312 MAX phase, and (c) FFT pattern of area *(v)* indicative of an $Al_3(Zr,Ti)_2$ intermetallic.

Fig. 8 is a high-magnification STEM micrograph corresponding to the $Al_2O_3$/film interface (area *(iii)*, Fig. 6a), and the FFT patterns of the encountered phases. Under the selected deposition conditions, substrate and thin film appear to have reacted through solid-state diffusion, probably due to the elevated temperature, as evidenced by the different chemistry and structure of the interface. The bright zone in Fig. 8a represents most of the film and corresponds to (Zr,Ti)C (Fig. 8b), while the darker zones correspond to the phases formed at the film/substrate interface. One may observe that the interface is not homogeneous, but rather comprises separate zones of the 211 MAX phase and intermetallic $Al_2(Zr,Ti)$, which were indexed from the FFT patterns in Fig. 8c and 8d, respectively. There are two possible scenarios that can explain such inhomogeneous interface: either the intermetallic is favoured because the lattice mismatch between $Al_2(Zr,Ti)$ and $Al_2O_3$ is smaller than the mismatch with the (Ti,Zr)C carbide and/or the 211 MAX phase (Fig. 8e), or the 211 MAX phase and the (Ti,Zr)C carbide, albeit more stable, cannot form due to high interfacial energies. However, the lattice parameter mismatch is more probable as both IMCs and carbides are known to be precursors of MAX phase compounds. [34] Indeed, it is likely that a reaction between $Al_2(Zr,Ti)$ and $(Ti_{0.5},Zr_{0.5})C$ has led to the formation of the 211 phase that is observed at the interface, causing discontinuities in the Al$_2$(Zr,Ti) layer and thickness variations. Furthermore, similar reactions occur within the thin film, since the 211 MAX phase was found invariably close to the Al$_3$(Zr,Ti)$_2$ intermetallic core, indicating that 211 phases tend to form close to an abundant Al source (see Fig. 7).

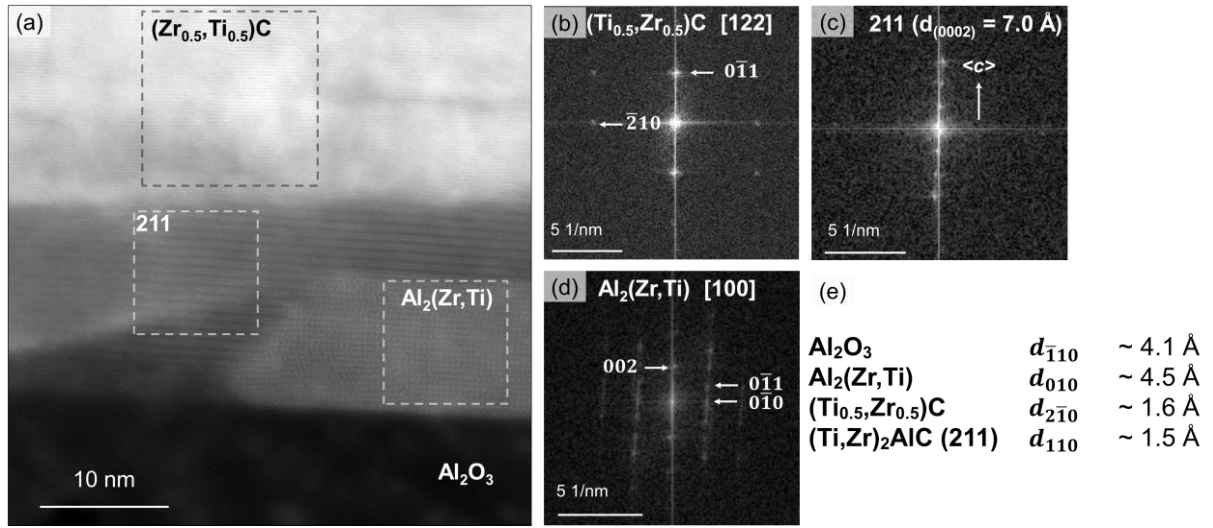

**Fig. 8.** (a) High magnification STEM micrograph of the thin film and the Al$_2$O$_3$/film interface. FFT patterns of the (b) zone denoted by (Ti,Zr)C, (c) zone denoted by 211, and (d) zone denoted by Al$_2$(Zr,Ti). (e) Interplanar spacing of each crystalline species parallel to the thin film surface.

In order to determine the Ti:Zr ratio of each MAX phase, several EDS spectra were collected from the numbered square areas shown in Fig. 9 and are summarized in Table 2. One may notice the variation in the Ti:Zr ratio, depending on the MAX phase compound. Indeed, the 312 MAX phase appears to be Ti-rich compared to the 211 phase, which maintains a Ti:Zr ratio close to 1:1. The (Ti,Zr)C and IMCs were all slightly more Zr-rich. The EDS data collected from the 211 phases indicate that the mismatch of the 211 MAX phase peaks in the XRD patterns of Fig. 4 and the reported (Ti$_{0.5}$,Zr$_{0.5}$)$_2$AlC peak positions can be largely attributed to internal stresses and oxygen content.

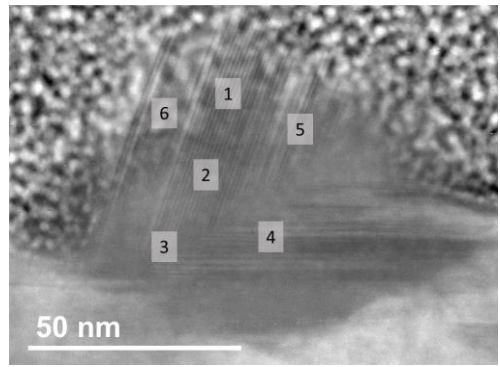

**Fig. 9.** STEM micrograph of a columnar protrusion and EDS point analyses.

**Table 2.** EDS point analyses of the MAX phase columnar protrusion in Fig. 9.

| Zone | MAX phase | Ti (at.%) | Zr (at.%) | Ti/(Ti+Zr) (%) | TM*/Al |
|---|---|---|---|---|---|
| 1 | 312 | 21.5 | 12.8 | 63 | 2.7 |
| 2 | 312 | 22.0 | 12.6 | 64 | 2.8 |
| 3 | 312 | 20.6 | 15.4 | 57 | 2.5 |
| 4 | 312/211 | 22.3 | 13.9 | 62 | 2.2 |
| 5 | 211 | 17.8 | 19.2 | 48 | 2.1 |
| 6 | 211 | 15.5 | 17.3 | 47 | 1.9 |

*TM: Transition Metal (Ti+Zr)

## 4. Conclusions

A $(Ti,Zr)_2AlC$ compound target was used to deposit a mixed $(Ti,Zr)C$ carbide containing up to 17 at.% of Al, as evidenced by ion beam analysis. High substrate temperatures and longer deposition times have resulted in films mostly composed of the $(Ti,Zr)C$ carbide with MAX phase protrusions. Both 312 and 211 MAX phases were identified with a Ti:Zr ratio of ~ 2:1 and 1:1, respectively. Moreover, two intermetallic phases were identified by means of XRD and STEM analyses, i.e., $Al_2(Zr,Ti)$ and $Al_3(Zr,Ti)_2$. These IMCs appear to be instrumental to the nucleation and growth of the MAX phases. Under the selected deposition conditions, the growth mechanism involved the deposition of an Al-containing $(Ti,Zr)C$ followed by the formation of IMCs, which in turn acted as Al sources for the growth of the MAX phases. Furthermore, the interface between the $Al_2O_3$ substrate and the film was inhomogeneous and mostly composed of the 211 MAX phase and $Al_2(Zr,Ti)$. The growth of the MAX phase was clearly affected by the substrate temperature. Indeed, up to 800 °C, only IMCs and (Zr,Ti)C are formed, indicating that the temperature must be above 900 °C for MAX phases formation.

These findings also support the solid state formation of MAX phases from the reaction of IMCs with the carbide. Finally, several challenges must be tackled before using such coatings to protect commercial zircaloy fuel cladding materials during high-temperature transients/accidents. First, the coating deposition temperature must be lowered so as to avoid degrading the properties of the substrate zircaloy clad. Second, the compatibility of such coatings with the coolant in all its states (water, steam), and the radiation tolerance of this coated fuel cladding material concept must be carefully assessed.

**Acknowledgements**

The authors acknowledge Dr. Thomas Lapauw for synthesizing the compound target. Support from the Swedish research council VR-RFI (#2017-00646_9) for the Accelerator based ion-technological centre, and from the Swedish Foundation for Strategic Research (contract RIF14-0053) for the tandem accelerator laboratory in Uppsala is gratefully acknowledged. This research has received funding from the Euratom research and training programme 2014–2018 under grant agreement No. 740415 (H2020 IL TROVATORE). We also acknowledge funding from Stiftelsen Olle Engkvist Byggmästare (grant no. 184-561), the Knut and Alice Wallenberg Foundation for Wallenberg Academy Fellowship grants and project funding (KAW 2015.0043), and the Swedish Government Strategic Research Area in Materials Science on Functional Materials at Linköping University (Faculty Grant SFO-Mat-LiU No. 2009 00971). The authors gratefully acknowledge the Hercules Foundation for Project AKUL/1319 (CombiS(T)EM)).

**References**

[1] A. Hoerling, J. Sjölén, H. Willmann, T. Larsson, M. Odén, L. Hultman, Thermal stability, microstructure and mechanical properties of $Ti_{1-x}Zr_xN$ thin films, Thin Solid Films. 516 (2008) 6421–6431. doi:10.1016/j.tsf.2007.12.133.

[2] V.V. Uglov, V.M. Anishchik, V.V. Khodasevich, Z.L. Prikhodko, S.V. Zlotski, G. Abadias, S.N. Dub, Structural characterization and mechanical properties of Ti-Zr-N coatings, deposited by vacuum arc, Surf. Coatings Technol. 180–181 (2004) 519–525.

doi:10.1016/j.surfcoat.2003.10.095.

[3]     T. Ma, P. Hedström, V. Ström, A. Masood, I. Borgh, A. Blomqvist, J. Odqvist, Self-organizing nanostructured lamellar (Ti,Zr)C - A superhard mixed carbide, Int. J. Refract. Met. Hard Mater. 51 (2015) 25–28. doi:10.1016/j.ijrmhm.2015.02.010.

[4]     E.A. Levashov, V. V. Kurbatkina, A.A. Zaitsev, S.I. Rupasov, E.I. Patsera, A.A. Chernyshev, Y. V. Zubavichus, A.A. Veligzhanin, Structure and properties of precipitation-hardening ceramic Ti-Zr-C and Ti-Ta-C materials, Phys. Met. Metallogr. 109 (2010) 95–105. doi:10.1134/S0031918X10010102.

[5]     M.W. Barsoum, MAX phases: Properties of machinable ternary carbides and nitrides, 2013. doi:10.1002/9783527654581.

[6]     M.W. Barsoum, $M_{n+1}AX_n$ phases: a new class of solids; thermodynamically stable nanolaminates, Prog. Solid State Chem. 28 (2000) 201–281. doi:10.1016/S0079-6786(00)00006-6.

[7]     P. Eklund, M. Beckers, U. Jansson, H. Högberg, L. Hultman, The $M_{n+1}AX_n$ phases: Materials science and thin-film processing, Thin Solid Films. 518 (2010) 1851–1878. doi:10.1016/j.tsf.2009.07.184.

[8]     M. Sokol, V. Natu, S. Kota, M.W. Barsoum, On the chemical diversity of the MAX phases, Trends Chem. 1 (2019) 210–223. doi:10.1016/j.trechm.2019.02.016.

[9]     M.W. Barsoum, T. El-Raghy, The MAX phases: Unique new carbide and nitride materials, Am. Sci. 89 (2001) 334–343. doi:10.1511/2001.4.334.

[10]    A. Abdulkadhim, M. to Baben, T. Takahashi, V. Schnabel, M. Hans, C. Polzer, P. Polcik, J.M. Schneider, Crystallization kinetics of amorphous $Cr_2AlC$ thin films, Surf. Coatings Technol. 206 (2011) 599–603. doi:10.1016/j.surfcoat.2011.06.003.

[11]    T. Cabioc'h, M. Alkazaz, M.F. Beaufort, J. Nicoläi, D. Eyidi, P. Eklund, $Ti_2AlN$ thin films synthesized by annealing of (Ti+Al)/AlN multilayers, Mater. Res. Bull. 80 (2016) 58–63. doi:10.1016/j.materresbull.2016.03.031.

[12]    A. Abdulkadhim, M. to Baben, T. Takahashi, V. Schnabel, M. Hans, C. Polzer, P. Polcik, J.M. Schneider, Crystallization kinetics of amorphous $Cr_2AlC$ thin films, Surf. Coatings Technol. (2011). doi:10.1016/j.surfcoat.2011.06.003.

[13]    Z. Feng, P. Ke, A. Wang, Preparation of $Ti_2AlC$ MAX phase coating by DC magnetron sputtering deposition and vacuum heat treatment, J. Mater. Sci. Technol. 31 (2015) 1193–1197. doi:10.1016/j.jmst.2015.10.014.

[14]    Z. Wang, W. Li, Y. Liu, J. Shuai, P. Ke, A. Wang, Diffusion-controlled intercalation approach to synthesize the $Ti_2AlC$ MAX phase coatings at low temperature of 550 °C, Appl. Surf. Sci. 502 (2020). doi:10.1016/j.apsusc.2019.144130.

[15]    J.J. Li, L.F. Hu, F.Z. Li, M.S. Li, Y.C. Zhou, Variation of microstructure and composition of the $Cr_2AlC$ coating prepared by sputtering at 370 and 500°C, Surf. Coatings Technol. 204 (2010) 3838–3845. doi:10.1016/j.surfcoat.2010.04.067.

[16]    R. Mertens, Z. Sun, D. Music, J.M. Schneider, Effect of the composition on the structure of Cr-Al-C investigated by combinatorial thin film synthesis and ab initio calculations,

Adv. Eng. Mater. 6 (2004) 903–907. doi:10.1002/adem.200400096.

[17] J. Nicolaï, C. Furgeaud, B.W. Fonrose, C. Bail, M.F. Beaufort, Formation mechanisms of Ti$_2$AlC MAX phase on SiC-4H using magnetron sputtering and post-annealing, Mater. Des. 144 (2018) 209–213. doi:10.1016/j.matdes.2018.02.046.

[18] J. Rosén, L. Ryves, P.O.Å. Persson, M.M.M. Bilek, Deposition of epitaxial Ti$_2$AlC thin films by pulsed cathodic arc, J. Appl. Phys. 101 (2007) 2006–2008. doi:10.1063/1.2709571.

[19] J. Frodelius, P. Eklund, M. Beckers, P.O.Å. Persson, H. Högberg, L. Hultman, Sputter deposition from a Ti$_2$AlC target: Process characterization and conditions for growth of Ti$_2$AlC, Thin Solid Films. 518 (2010) 1621–1626. doi:10.1016/j.tsf.2009.11.059.

[20] O. Wilhelmsson, J.P. Palmquist, E. Lewin, J. Emmerlich, P. Eklund, P.O.Å. Persson, H. Högberg, S. Li, R. Ahuja, O. Eriksson, L. Hultman, U. Jansson, Deposition and characterization of ternary thin films within the Ti-Al-C system by DC magnetron sputtering, J. Cryst. Growth. 291 (2006) 290–300. doi:10.1016/j.jcrysgro.2006.03.008.

[21] T. Lapauw, D. Tytko, K. Vanmeensel, S. Huang, P.-P. Choi, D. Raabe, E.N. Caspi, O. Ozeri, M. to Baben, J.M. Schneider, K. Lambrinou, J. Vleugels, (Nb$_x$, Zr$_{1-x}$)$_4$AlC$_3$ MAX phase solid solutions: Processing, mechanical properties, and density functional theory calculations, Inorg. Chem. 55 (2016) 5445–5452. doi:10.1021/acs.inorgchem.6b00484.

[22] D. Horlait, S. Grasso, A. Chroneos, W.E. Lee, Attempts to synthesise quaternary MAX phases (Zr,M)$_2$AlC and Zr$_2$(Al,A)C as a way to approach Zr$_2$AlC, Mater. Res. Lett. 4 (2016) 137–144. doi:10.1080/21663831.2016.1143053.

[23] M.A. Ali, M.M. Hossain, N. Jahan, A.K.M.A. Islam, S.H. Naqib, Newly synthesized Zr$_2$AlC, Zr$_2$(Al$_{0.58}$,Bi$_{0.42}$)C, Zr$_2$(Al$_{0.2}$,Sn$_{0.8}$)C, and Zr$_2$(Al$_{0.3}$,Sb$_{0.7}$)C MAX phases: A DFT based first-principles study, Comput. Mater. Sci. 131 (2017) 139–145. doi:10.1016/j.commatsci.2017.01.048.

[24] W. Yu, V. Mauchamp, T. Cabioc'h, D. Magne, L. Gence, L. Piraux, Solid solution effects in the Ti$_2$Al(C$_x$,N$_y$) MAX phases: Synthesis, microstructure, electronic structure and transport properties, Acta Mater. 80 (2014) 421–434. doi:10.1016/j.actamat.2014.07.064.

[25] J. Halim, P. Chartier, T. Basyuk, T. Prikhna, E.N. Caspi, M.W. Barsoum, T. Cabioc'h, Structure and thermal expansion of (Cr$_x$,V$_{1-x}$)$_{n+1}$AlC$_n$ phases measured by X-ray diffraction, J. Eur. Ceram. Soc. 37 (2017) 15–21. doi:10.1016/j.jeurceramsoc.2016.07.022.

[26] A. Petruhins, A.S. Ingason, J. Lu, F. Magnus, S. Olafsson, J. Rosén, Synthesis and characterization of magnetic (Cr$_{0.5}$,Mn$_{0.5}$)$_2$GaC thin films, J. Mater. Sci. 50 (2015) 4495–4502. doi:10.1007/s10853-015-8999-8.

[27] E. Zapata-Solvas, M.A. Hadi, D. Horlait, D.C. Parfitt, A. Thibaud, A. Chroneos, W.E. Lee, Synthesis and physical properties of (Zr$_{1-x}$,Ti$_x$)$_3$AlC$_2$ MAX phases, J. Am. Ceram. Soc. (2017) 3393–3401. doi:10.1111/jace.14870.

[28] B. Tunca, T. Lapauw, O.M. Karakulina, M. Batuk, T. Cabioc'h, J. Hadermann, R. Delville, K. Lambrinou, J. Vleugels, Synthesis of MAX phases in the Zr-Ti-Al-C system,

Inorg. Chem. 56 (2017) 3489–3498. doi:10.1021/acs.inorgchem.6b03057.

[29] T. Cabioc'h, P. Eklund, V. Mauchamp, M. Jaouen, M.W. Barsoum, Tailoring of the thermal expansion of $Cr_2(Al_x,Ge_{1-x})C$ phases, J. Eur. Ceram. Soc. 4 (2013) 897–904.

[30] R. Meshkian, A.S. Ingason, M. Dahlqvist, A. Petruhins, U.B. Arnalds, F. Magnus, J. Lu, J. Rosén, Theoretical stability, thin film synthesis and transport properties of the $Mo_{n+1}GaC_n$ MAX phase, Phys. Status Solidi - Rapid Res. Lett. 9 (2015) 197–201. doi:10.1002/pssr.201409543.

[31] P. Eklund, M. Beckers, J. Frodelius, H. Högberg, L. Hultman, Magnetron sputtering of $Ti_3SiC_2$ thin films from a compound target, J. Vac. Sci. Technol. A Vacuum, Surfaces, Film. 25 (2007) 1381. doi:10.1116/1.2757178.

[32] J.P. Palmquist, U. Jansson, T. Seppänen, P.O.Å. Persson, J. Birch, L. Hultman, P. Isberg, Magnetron sputtered epitaxial single-phase $Ti_3SiC_2$ thin films, Appl. Phys. Lett. 81 (2002) 835–837. doi:10.1063/1.1494865.

[33] C. Walter, D.P. Sigumonrong, T. El-Raghy, J.M. Schneider, Towards large area deposition of $Cr_2AlC$ on steel, Thin Solid Films. 515 (2006) 389–393. doi:10.1016/j.tsf.2005.12.219.

[34] T. Lapauw, K. Lambrinou, T. Cabioc'h, J. Halim, J. Lu, A. Pesach, O. Rivin, O. Ozeri, E.N. Caspi, L. Hultman, P. Eklund, J. Rosén, M.W. Barsoum, J. Vleugels, Synthesis of the new MAX phase $Zr_2AlC$, J. Eur. Ceram. Soc. (2016) 1847–1853.

[35] T. Lapauw, J. Halim, J. Lu, T. Cabioc'h, L. Hultman, M.W. Barsoum, K. Lambrinou, J. Vleugels, Synthesis of the novel $Zr_3AlC_2$ MAX phase, J. Eur. Ceram. Soc. 36 (2016) 943–947. doi:10.1016/j.jeurceramsoc.2015.10.011.

[36] C.C. Lai, M.D. Tucker, J. Lu, J. Jensen, G. Greczynski, P. Eklund, J. Rosén, Synthesis and characterization of $Zr_2Al_3C_4$ thin films, Thin Solid Films. 595 (2015) 142–147. doi:10.1016/j.tsf.2015.10.079.

[37] R. Zhang, G. Chen, W. Han, Synthesis, mechanical and physical properties of bulk $Zr_2Al_4C_5$ ceramic, Mater. Chem. Phys. 119 (2010) 261–265. doi:10.1016/j.matchemphys.2009.08.051.

[38] K. Arstila, J. Julin, M.I. Laitinen, J. Aalto, T. Konu, S. Kärkkäinen, S. Rahkonen, M. Raunio, J. Itkonen, J.P. Santanen, T. Tuovinen, T. Sajavaara, Potku, New analysis software for heavy ion elastic recoil detection analysis, Nucl. Instruments Methods Phys. Res. Sect. B Beam Interact. with Mater. Atoms. 331 (2014) 34–41. doi:10.1016/j.nimb.2014.02.016.

[39] M. Magnuson, L. Tengdelius, G. Greczynski, F. Eriksson, J. Jensen, J. Lu, M. Samuelsson, P. Eklund, L. Hultman, H. Högberg, Compositional dependence of epitaxial $Ti_{n+1}SiC_n$ MAX-phase thin films grown from a $Ti_3SiC_2$ compound target, J. Vac. Sci. Technol. A. 37 (2019) 021506. doi:10.1116/1.5065468.

[40] P.O.Å. Persson, C. Höglund, J. Birch, L. Hultman, $Ti_2Al(O,N)$ formation by solid-state reaction between substoichiometric TiN thin films and $Al_2O_3(0001)$ substrates, Thin Solid Films. 519 (2011) 2421–2425. doi:10.1016/j.tsf.2010.12.002.

[41] P.O.Å. Persson, J. Rosén, D.R. McKenzie, M.M.M. Bilek, C. Höglund, A solid phase

reaction between TiC$_x$ thin films and Al$_2$O$_3$ substrates, J. Appl. Phys. 103 (2008) 1–4. doi:10.1063/1.2896637.